\documentclass[11pt]{article}
\pdfoutput=1
\usepackage[a4paper, margin=0.9in]{geometry}
\usepackage{authblk}
\usepackage{graphicx}
\usepackage{dcolumn}
\usepackage{amsmath}
\usepackage{amssymb}
\usepackage{textcomp}
\usepackage{verbatim}
\usepackage{float}
\usepackage[symbol]{footmisc}
\setfnsymbol{wiley}
\usepackage{hyperref}
\hypersetup{
    colorlinks=true,
    linkcolor=blue,
    filecolor=blue,
    urlcolor=blue,
    citecolor = blue,
    pdfauthor=author,
}
\usepackage{multirow}
\usepackage[T1]{fontenc}
\usepackage[normalem]{ulem}
\usepackage[square,sort,comma,numbers]{natbib}

\begin{document}
\title{Probing Gigahertz Coherent Acoustic Phonons in TiO$_{2}$ Mesoporous Thin Films}

\author[1]{E. R. Cardozo de Oliveira\footnote{These authors contributed equally to this work}$^{,}$}

\author[1]{C. Xiang$^{*,}$}

\author[1,2]{M. Esmann}

\author[3]{N. Lopez Abdala}

\author[4]{M. C. Fuertes}

\author[5]{A. Bruchhausen}

\author[5]{H. Pastoriza}

\author[6]{B. Perrin}

\author[3]{G. J. A. A. Soler-Illia\footnote{Corresponding authors: daniel.kimura@c2n.upsaclay.fr, gsoler-illia@unsam.edu.ar}$^{,}$}

\author[1]{N. D. Lanzillotti-Kimura$^{**,}$}

\affil[1]{Université Paris-Saclay, CNRS, Centre de Nanosciences et de Nanotechnologies, 91120 Palaiseau, France}
\affil[2]{Institute for Physics, University of Oldenburg, 26129 Oldenburg, Germany}
\affil[3]{Instituto de NanoSistemas - Universidad Nacional de San Martín-CONICET, Buenos Aires, Argentina}
\affil[4]{Gerencia Química, Inst. de Nanociencia y Nanotecnología, CNEA-CONICET, Buenos Aires, Argentina}
\affil[5]{Centro Atómico Bariloche, Inst. de Nanociencia y Nanotecnología, CNEA-CONICET, Rio Negro, Argentina}
\affil[6]{Sorbonne Université, CNRS, Institut des NanoSciences de Paris, INSP, F-75005 Paris, France}

\date{}
\maketitle
\begin{abstract}

Ultrahigh-frequency acoustic-phonon resonators usually require atomically flat interfaces to avoid phonon scattering and dephasing, leading to expensive fabrication processes, such as molecular beam epitaxy. In contrast, mesoporous thin films are based on inexpensive wet chemical fabrication techniques. Here, we report mesoporous titanium dioxide-based acoustic resonators with resonances up to 90 GHz, and quality factors from 3 to 7. Numerical simulations show a good agreement with the picosecond ultrasonics experiments. We also numerically study the effect of changes in the speed of sound on the performance of the resonator. This change could be induced by liquid infiltration into the mesopores. Our findings constitute the first step towards the engineering of building blocks based on mesoporous thin films for reconfigurable optoacoustic sensors.

\end{abstract}

\section{Introduction}

Coherent acoustic phonons, with frequencies in the GHz - THz range, have associated wavelengths between a few and hundreds of nanometers.~\cite{Trigo2002,balandinNanophononicsPhononEngineering2005,rozas_lifetime_2009,beardsley_coherent_2010,Volz2016,della_picca_tailored_2016,lamberti_optomechanical_2017,de_luca_phonon_2019,esmann_brillouin_2019} Among other applications, they are suitable for high resolution nanoimaging, non-destructive testing, and sensing. Acoustic phonon dynamics have been explored in systems such as plasmonic nanostructures,~\cite{arbouet_optical_2006,obrien_ultrafast_2014,guillet_ultrafast_2019,poblet_acoustic_2021} metasurfaces,~\cite{lanzillotti-kimura_polarization-controlled_2018} oxides,~\cite{soukiassian_acoustic_2007,lanzillotti-kimura_enhancement_2010,vasileiadis_progress_2021,cang_fundamentals_2022} and semiconductor heterostructures~\cite{lanzillotti-kimura_resonant_2009,anguianoMicropillarResonatorsOptomechanics2017,chafatinos_polariton-driven_2020,ortiz_topological_2021} with layer thicknesses on the nanometric scale.~\cite{arregui_coherent_2019,lanzillotti-kimura_resonant_2009,lanzillotti-kimura_enhanced_2011} To precisely tailor the nanophononic response and obtain high quality devices, expensive and complex growth and processing techniques are usually employed, including molecular beam epitaxy and electron beam lithography. 

Conversely, mesoporous structures rely on cheap and reproducible bottom-up fabrication processes derived from soft chemistry, carried out in mild conditions,~\cite{soler-illia_mesoporous_2006,gazoni_designed_2017} and are able to sustain gigahertz acoustic resonances.~\cite{gomopoulos_one-dimensional_2010,Benetti2018,Abdala2020} Brillouin light scattering experiments have been employed to demonstrate phononic band gaps in the 10 - 20 GHz range in periodic stacks of alternating layers of porous silicon dioxide (SiO$_{2}$) and poly- (methyl methacrylate) (PMMA). \cite{gomopoulos_one-dimensional_2010,schneider_engineering_2012,schneider_defect-controlled_2013} More recently, mesoporous thin films (MTFs) based on SiO$_{2}$ have been shown to support coherent acoustic modes between 5 and 100 GHz, with Q-factors ranging from 5 to 17.~\cite{Abdala2020}. MTFs are also suitable for photonic sensing applications due to the high surface-to-volume ratio and tailorable mesopores.~\cite{auguie_tamm_2014,fuertes_photonic_2007} By liquid infiltration into the nanopores, a modulation of the optical and elastic properties of the material could be achieved, also enabling chemical functionalization in nanoacoustic devices.~\cite{Benetti2018, thelen_laser-excited_2021} 

A well-established material, with numerous applications for both the dense and mesoporous forms, is titanium dioxide (TiO$_{2}$). Applications include photocatalysis,~\cite{suarez_photocatalytic_2011,hussein_mesoporous_2013} implants,~\cite{harmankaya_raloxifene_2013} photovoltaics,~\cite{violi_highly_2012} energy harvesting and storage, and sensing.~\cite{pan_block_2011} These materials are promising candidates for nanoacoustics. For instance, TiO$_{2}$ anatase nanoparticles presented a particularly strong induced exciton shift when applying an acoustic strain pulse.~\cite{baldini_phonon-driven_2018,baldini_exciton_2019}

In this work, we employ coherent phonon generation and detection techniques to study acoustic resonators based on TiO$_{2}$ MTFs, with resonances up to 90 GHz. Our results indicate that GHz acoustic resonators based on mesoporous structures are not limited to SiO$_{2}$, and open the possibility to more complex heterostructures formed by different materials. In addition, we theoretically investigate the effect of changes in the elastic properties of the mesoporous matrix on the acoustic resonances, which can be achieved via liquid infiltration into the pores. This platform constitutes a promising building block for developing environment-responsive nanosystems for nanoacoustic sensing and reconfigurable optoacoustic nanodevices based on soft and inexpensive fabrication methods.

\section{Experiments and Simulations}

\subsection{Sample Fabrication}

The TiO$_{2}$ thin films are synthesized using the sol-gel method. The ordered mesoporosity is obtained using the evaporation-induced self-assembly of surfactants.~\cite{soler-illia_mesoporous_2006} The fabrication details can be found in reference~\citenum{Abdala2020} for silica MTFs, in which the same approach is employed. Ethanolic solutions of TiCl$_{4}$ are first prepared, and then the water is added; to obtain the porosity, the surfactant Pluronic F127 is finally added to the precursor solution. The final molar ratio of the sols is TiCl$_{4}$:H$_{2}$O:EtOH 1:10:40 to obtain the dense films and TiCl$_{4}$:F127:H$_{2}$O:EtOH 1:0.005:10:40 to synthesize the mesoporous materials. These sols are used immediately after preparation to deposit the films on top of clean Si substrates by dip coating.

\begin{figure}
	\par
	\begin{center}
		\includegraphics[scale=0.42]{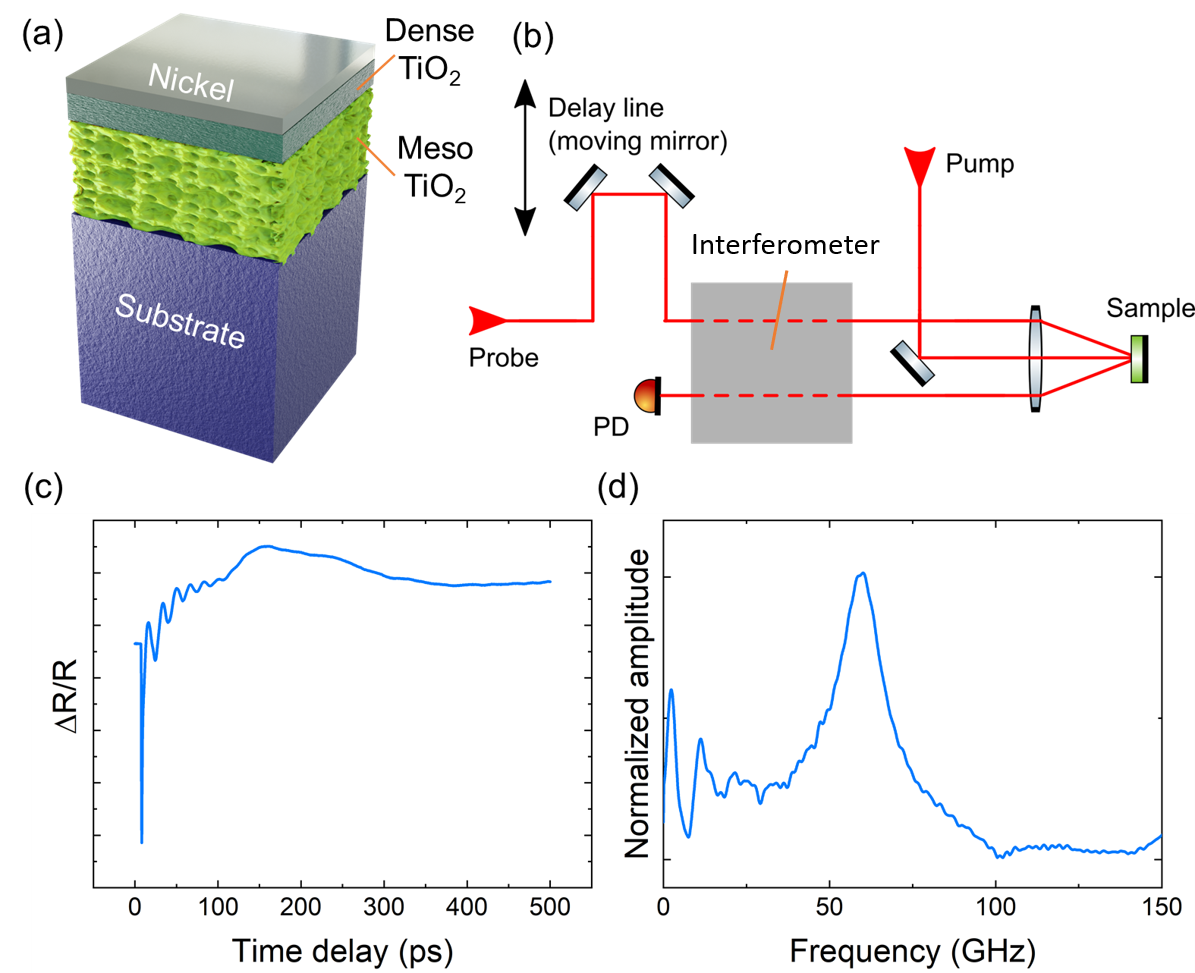} 
	\end{center}
	\par
	\vspace{-0.6cm} \caption{ (a) Schematics of the mesoporous TiO$_{2}$ sample design. (b) Experimental setup of the reflectometric pump-probe setup. An interferometric measurement can also be implemented, indicated by the gray box.~\cite{perrin_interferometric_1999,matsuda_reflection_2002-1} (c) Reflectivity timetrace of the mesoporous TiO$_{2}$ sample C. (d) Phononic spectrum corresponding to the timetrace shown in (c).}
	\label{sample}
\end{figure}

The schematics of the studied structure is shown in Fig~\ref{sample}(a). First, a mesostructured titania film is deposited on the substrate and submitted to a stabilization process up to 200ºC to consolidate the oxide and stabilize the mesostructure.~\cite{angelome_multifunctional_2006} Then, a capping layer of dense titania is synthesized and stabilized with the same treatment. Afterwards, the bilayer is submitted to a final calcination step of 2 h at 350°C with a 1°C/min heating ramp; this treatment is performed to eliminate the pores template and generate an accessible and interconnected mesoporosity.~\cite{fuertes_sorption_2008} Besides, it leads to a phase transition from amorphous titania to anatase.~\cite{soler-illia_critical_2012} After these steps, homogeneous and crack-free thin films of cm dimensions are achieved.

Afterwards, a $\sim$32 nm-thick nickel cover layer is deposited by vacuum thermal evaporation using a homemade system. The dense layer is necessary to avoid nickel diffusion through the mesopores, and the Ni film acts as the acousto-optical transducer for coherent phonon generation and detection.~\cite{huynh_subterahertz_2006}

\begin{table}
\caption{Structural details of the layers from samples A, B, C and Control. Best fit (nominal) thickness values of the dense TiO$_{2}$, nickel and mesoporous layers, and the porosity.}
\begin{center}
\begin{tabular}{c|ccclcl}
\hline
\multirow{5}{*}{\textbf{Sample}} & \multicolumn{2}{c}{\multirow{2}{*}{\textbf{\begin{tabular}[c]{@{}c@{}}Mesoporous \\ TiO$_{2}$\end{tabular}}}} & \multicolumn{2}{c}{\multirow{2}{*}{\textbf{\begin{tabular}[c]{@{}c@{}}Dense \\ TiO$_{2}$\end{tabular}}}} & \multicolumn{2}{c}{\multirow{2}{*}{\textbf{Ni}}} \\
& \multicolumn{2}{c}{} & \multicolumn{2}{c}{} & \multicolumn{2}{c}{} \\ \cline{2-7} 
 & \multirow{3}{*}{\textbf{\begin{tabular}[c]{@{}c@{}}Thickness \\ (nm)\end{tabular}}} & \multicolumn{1}{c|}{\multirow{3}{*}{\textbf{\begin{tabular}[c]{@{}c@{}}Porosity \\ (\%)\end{tabular}}}} & \multicolumn{4}{c}{\multirow{3}{*}{\textbf{Thickness (nm)}}} \\
 & & \multicolumn{1}{c|}{} & \multicolumn{4}{c}{} \\
& & \multicolumn{1}{c|}{} & \multicolumn{4}{c}{} \\
\multirow{2}{*}{A} & \multirow{2}{*}{110 (103)} & \multicolumn{1}{c|}{\multirow{2}{*}{48}} & \multicolumn{2}{c}{\multirow{2}{*}{18 (14)}} & \multicolumn{2}{c}{\multirow{2}{*}{35 (32)}} \\
& & \multicolumn{1}{c|}{} & \multicolumn{2}{c}{} & \multicolumn{2}{c}{} \\
\multirow{2}{*}{B} & \multirow{2}{*}{160 (144)} & \multicolumn{1}{c|}{\multirow{2}{*}{40}} & \multicolumn{2}{c}{\multirow{2}{*}{30 (26)}} & \multicolumn{2}{c}{\multirow{2}{*}{31 (32)}} \\
& & \multicolumn{1}{c|}{} & \multicolumn{2}{c}{} & \multicolumn{2}{c}{} \\
\multirow{2}{*}{C} & \multirow{2}{*}{185 (191)} & \multicolumn{1}{c|}{\multirow{2}{*}{44}} & \multicolumn{2}{c}{\multirow{2}{*}{32 (25)}} & \multicolumn{2}{c}{\multirow{2}{*}{45 (32)}} \\
& & \multicolumn{1}{c|}{} & \multicolumn{2}{c}{} & \multicolumn{2}{c}{} \\
\multirow{2}{*}{Control} & \multirow{2}{*}{--} & \multicolumn{1}{c|}{\multirow{2}{*}{--}} & \multicolumn{2}{c}{\multirow{2}{*}{68 (60)}} & \multicolumn{2}{c}{\multirow{2}{*}{29.5 (32)}} \\
& & \multicolumn{1}{c|}{} & \multicolumn{2}{c}{} & \multicolumn{2}{c}{} \\ \hline
\end{tabular}
\end{center}
\label{table_1}
\end{table}

Four samples are fabricated: three structures designed with different mesoporous layer thicknesses, defined by the dip-coating withdrawal speeds of 1, 2 and 3 mm/s, and one control sample without the mesoporous thin film. The dense layer withdrawal speed is set constant to 0.2 mm/s. The structures are respectively labeled as A, B, C, and Control. Table~\ref{table_1} presents the thickness and porosity parameters, derived from ellipsometry measurements (SOPRA GES5E ellipsometer).~\cite{boissiere_porosity_2005}

\subsection{Coherent phonon generation and detection}

The coherent acoustic phonon dynamics is studied via time-domain Brillouin scattering (TDBS)~\cite{ruello_physical_2015} in a pump-probe setup (see Fig.~\ref{sample}(b)). An incident pulsed pump laser ($\lambda =$ 758 nm, 200 fs pulse duration, 80 MHz repetition rate, 275 mW equivalent CW power) is partially absorbed by the Ni transducer and excites acoustic phonons via photoinduced stress processes.~\cite{ruello_physical_2015} A second ultrafast laser pulse, namely the probe (typically 5 mW power), delayed with respect to the pump, detects the instantaneous optical reflectivity modulated by the coherent acoustic phonons. Two different processes drive these modulations: the photoelastic interaction, i.e., the changes in the optical properties of the structure due to strain; and the surface displacement induced by the presence of phonons. For the first process, the coherent acoustic phonons induce a change in the index of refraction in the structure proportional to the strain, which will then modulate the reflectivity of the probe. For the latter process, surface displacement detection is performed by implementing a Sagnac interferometer.~\cite{perrin_interferometric_1999,matsuda_reflection_2002-1} Further experimental details can be found in Ref. \citenum{Abdala2020}. The reflectivity setup is simpler in terms of optical alignment. However, depending on the studied materials and layer ordering, the modulation of the photoelastic properties is weak and the detection of coherent acoustic phonons becomes impossible. In such cases, the interferometric setup must be employed.~\cite{perrin_interferometric_1999}

Figure~\ref{sample}(c) depicts a typical transient reflectivity timetrace in the interferometric configuration for a TiO$_{2}$-based mesoporous sample.  Coherent oscillations are visible for time delay <120 ps. They result from longitudinal coherent phonons modulating the optical properties of the structure. By performing a Fourier transform, we extract the phononic response in the frequency domain (Fig~\ref{sample}(d)).

\subsection{Numerical Simulation}

To investigate the acoustic resonances, we simulate the acoustic strain, displacement, and electric fields by implementing a transfer matrix method to solve the respective wave equations for a multilayered structure with contrasting refractive indices and acoustic impedances. \cite{thomsen_surface_1986,Fainstein2007,lanzillotti-kimura_theory_2011-1,lanzillotti-kimura_coherent_2011-1} By calculating the normalized optical and acoustic solutions we can simulate the phonon generation spectrum by integrating the strain, the electric field square modulus and the photoelastic constant. The phonon detection is then obtained by an overlap integral of the generation spectrum with the photoelastic constant, the strain, and the electric field square. \cite{Fainstein2007,lanzillotti-kimura_theory_2011-1,lanzillotti-kimura_coherent_2011-1} The material parameters used in the simulations are shown in Table~\ref{table_2}. The surface displacement is also simulated by taking the product of the solutions of the phonon displacement amplitude at the interface between the air and the last layer of the structure, with the phonon generation spectrum. It is assumed that the optical absorption and coherent phonon generation are entirely limited to the nickel transducer layer.

\begin{table}
	\caption{Optical and elastic properties of the studied materials for the numerical simulation. For the mesoporous TiO$_{2}$, the index of refraction is derived from the Bruggemann approximation for a mixture of two dielectric media states~\cite{fungApplicationValidityEffective2019}: dense TiO$_{2}$ matrix and the air (pores), according to the porosity on table~\ref{table_1}.}
  \begin{center}
	\begin{tabular}{cccc}
		\hline
		\multirow{3}{*}{\textbf{Material}} & \multirow{3}{*}{\textbf{\begin{tabular}[c]{@{}c@{}}Index of \\ refraction\end{tabular}}} & \multirow{3}{*}{\textbf{\begin{tabular}[c]{@{}c@{}}Speed of \\ sound (m/s)\end{tabular}}} & \multirow{3}{*}{\textbf{\begin{tabular}[c]{@{}c@{}}Density\\  (g/cm$^{3}$)\end{tabular}}} \\
		& & & \\ & & & \\ \hline
		\multirow{2}{*}{TiO$_{2}$} & \multirow{2}{*}{2.56} & \multirow{2}{*}{6700} & \multirow{2}{*}{2.9} \\	& & & \\
		\multirow{2}{*}{SiO$_{2}$} & \multirow{2}{*}{1.5375} & \multirow{2}{*}{5750} & \multirow{2}{*}{2.2} \\ & & & \\
		\multirow{2}{*}{Air} & \multirow{2}{*}{1.00028} & \multirow{2}{*}{343} & \multirow{2}{*}{1.275E-3} \\ & & & \\
		\multirow{2}{*}{Nickel} & \multirow{2}{*}{2.218+4.893i} & \multirow{2}{*}{4970} & \multirow{2}{*}{8.908} \\ & & & \\
		\multirow{2}{*}{\begin{tabular}[c]{@{}c@{}}TiO$_{2}$ \\ (mesoporous)\end{tabular}} & \multirow{2}{*}{(1.76 -- 1.90)~\cite{fungApplicationValidityEffective2019}} & \multirow{2}{*}{\begin{tabular}[c]{@{}c@{}}$4355 $\end{tabular}}             & \multirow{2}{*}{\begin{tabular}[c]{@{}c@{}}$2.9$\end{tabular}} \\ & & & \\ \hline
	\end{tabular}
 \end{center}
	\label{table_2}
\end{table}

\section{Results and Discussion}

\subsection{Surface displacement and photoelastic interaction}

\begin{figure}
	\par
	\begin{center}
		\includegraphics[scale=0.43]{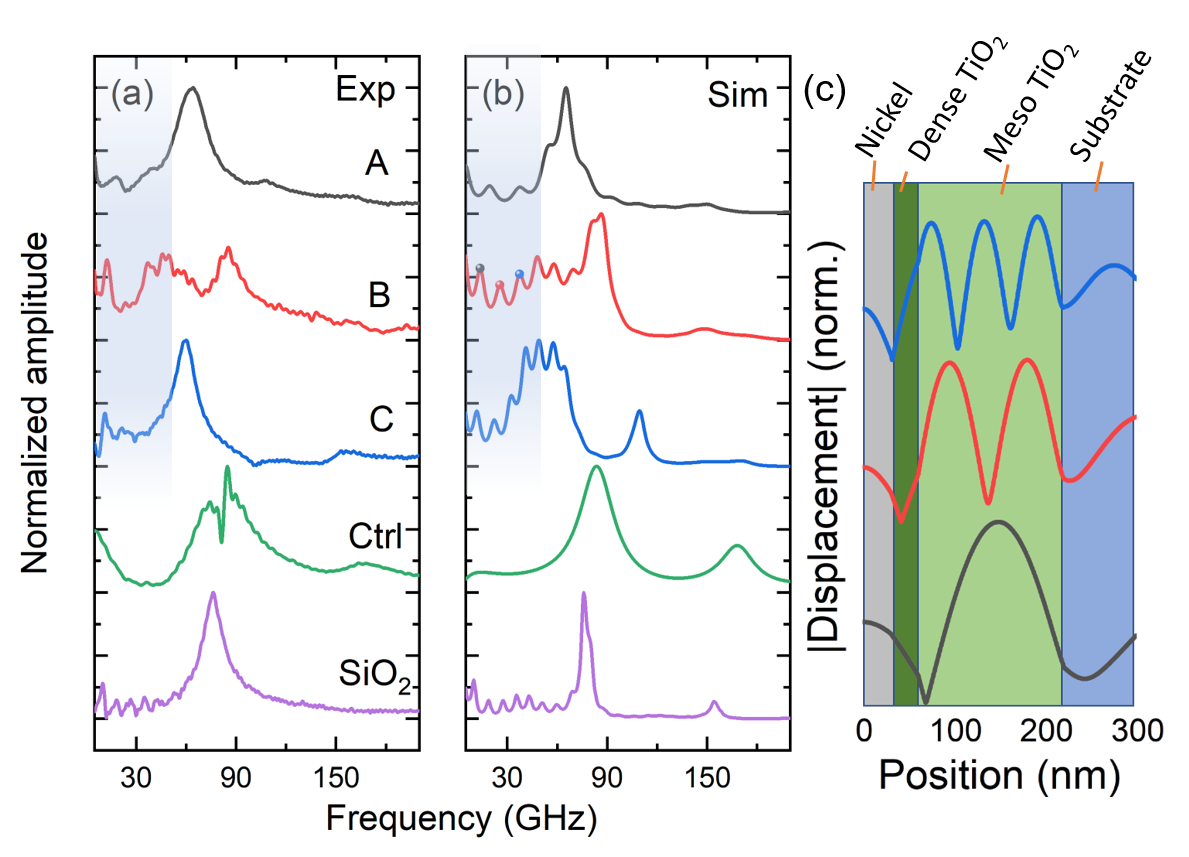} 
	\end{center}
	\par
	\vspace{-0.6cm} \caption{ (a) Experimental and (b) simulated spectra of the surface displacement of TiO$_{2}$-based samples A, B, C and Control, and SiO$_{2}$-based sample C (Ref. \citenum{Abdala2020}), for comparison. The low-frequency-shaded areas (up to 50 GHz) indicate the first confined modes in the mesoporous layer.  (c) Displacement field of the modes at 13, 25.5 and 38 GHz on TiO$_{2}$ sample B, respectively indicated as black, red and blue circles in (b).}
	\label{fft}
\end{figure}

The experimental surface displacement spectra of the TiO$_{2}$-based mesoporous samples A, B and C, and the control sample without mesoporous layer, are presented in Fig~\ref{fft}(a). An intense and broad peak is present between 55 GHz and 90 GHz. This peak is associated to an acoustic resonance in the nickel and dense TiO$_{2}$ bilayer.~\cite{Abdala2020} At lower frequencies, low amplitude peaks are resolved, which correspond to coherent acoustic phonons confined in the mesoporous layer. For comparison, the results for the mesoporous SiO$_{2}$ sample, reported in reference ~\citenum{Abdala2020}, are also displayed in Fig~\ref{fft}(a,b), where the acoustic modes at the mesoporous layer are better resolved. In contrast, the Control structure, without a mesoporous layer, does not show any resonances below 40 GHz, as expected. The spectrum associated to this sample presents a peak at 84.8 GHz, corresponding to the resonance on metallic and dense layers, with a dip at 81.2 GHz, probably originating from a destructive interference between the surface displacement and Brillouin scattering contributions of the signal.

The simulation of the surface displacements are shown in Fig.~\ref{fft}(b). A good agreement between the experimental data and simulation results is achieved. The interference dip present in the Control sample is not reproduced in the simulation as Brillouin scattering is not considered simultaneously with the surface displacement effect. Furthermore, the experimental results of sample C show a dip at $\sim$ 100 GHz, whereas a peak at the same frequency is present in the simulation.

\begin{figure}
	\par
	\begin{center}
		\includegraphics[scale=0.58]{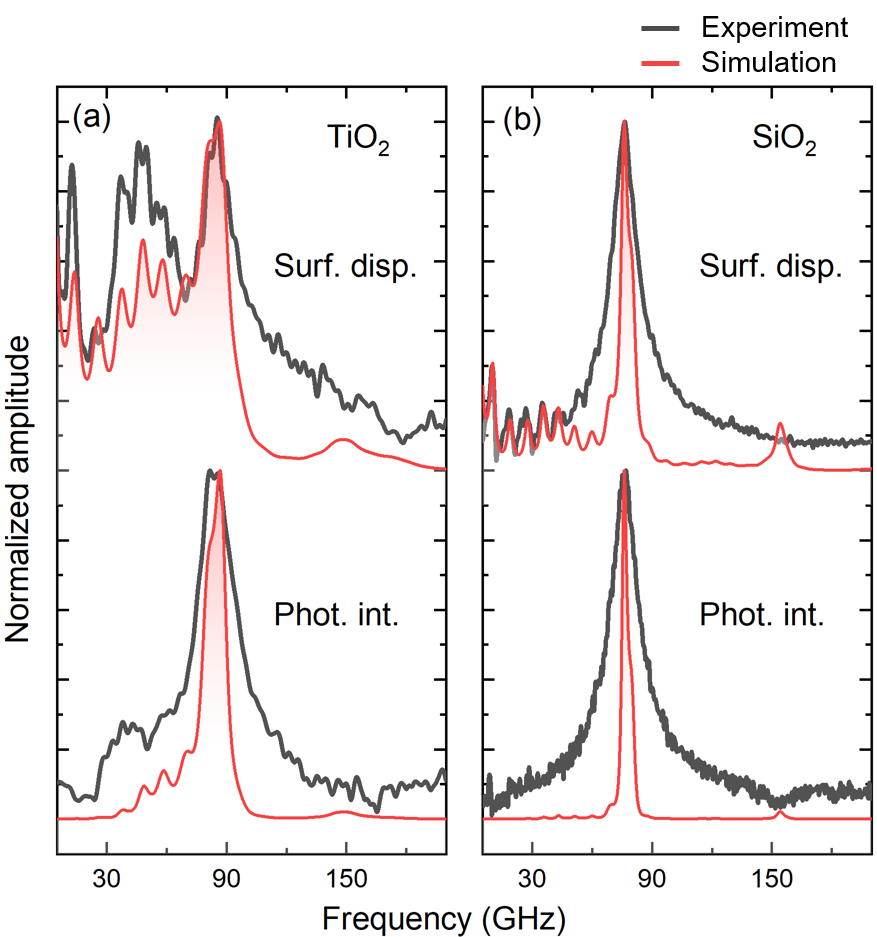} 
	\end{center}
	\par
	\vspace{-0.5cm} \caption{Experimental results (black line) and the respective TMM simulations (red area) on surface displacement and photoelastic interaction for (a) TiO$_{2}$ sample B and (b) SiO$_{2}$ sample C, from reference \citenum{Abdala2020}. Acoustic modes at the low-frequency region are only present at the surface displacement spectra for both TiO$_{2}$ and SiO$_{2}$.}
	\label{int-ref}
\end{figure}

In order to get a finer picture of the phononic behavior of these structures, it is worth comparing the signals -measured and calculated- corresponding to the surface displacement (interferometry) and modulation of the index of refraction (reflectometry, photoelastic effect). Figures~\ref{int-ref}(a) and (b) display the phononic spectra of both modulation effects for the TiO$_{2}$ (sample B) and SiO$_{2}$ structures, respectively. For both materials, the intense peak is present in the two cases. However, the lower frequency confined modes are observed uniquely in the interferometric dataset, in which the first five acoustic resonances at 12.6, 23.9, 36.9, 47.7 and 57.9 GHz have respective quality factors of 2.7, 4.7, 4.1, 3.9 and 6.9, within a 10$\%$ error.

Simulations for the surface displacement and the photoelastic interaction for both structures are displayed as red-shadowed lines in Figs.~\ref{int-ref}(a) and (b). They reproduce the main peaks of the experimental datasets for both cases. Furthermore, the low-frequency modes in the photoelastic interaction calculation, associated to the mesoporous resonances, are negligible when compared to the interferometric simulation, in accordance with the experimental results.

The lack of a photoelastic interaction contribution in both experiment and simulation supports the hypothesis that the respective vibrational resonances are confined in the mesoporous layer. Note that the mesoporous material is transparent for the incident laser wavelength -with an associated weak photoelastic constant-. For the surface displacement results, the acoustic modes in the soft mesoporous layer leads the whole structure to vibrate, thus, contributing to the interferometric detection. 

\subsection{Surface displacement dependence on TiO$_{2}$ speed of sound}

Elastic properties of TiO$_{2}$ thin films, such as mass density ($\rho$), Young's modulus ($Y$) and speed of sound ($v$) depend on several factors such as fabrication method, layer thickness, annealing temperature and crystallinity. This dependence leads to a broad range of reported values for $\rho$ (from 2.9 to 3.9 g/cm$^{3}$) and $Y$ (from 80 to 250 GPa), for TiO$_{2}$ in its anatase phase.~\cite{pillai_sol-gel_2017, yang_compositional_2005,ottermann_youngs_1996,ferrara_hydrophilic_2010,olofinjana_evaluation_2000,bendavid_deposition_2000,soares_hardness_2008,borgese_young_2012} The speed of propagation of longitudinal acoustic modes is calculated according to:
\begin{equation}
    v = \sqrt{\frac{Y(1-\nu)}{\rho(1+\nu)(1-2\nu)}},
\end{equation}
where $\nu$ is the Poisson ratio. Considering a Poisson ratio of 0.27 for TiO$_{2}$, the broad range of values for Y and $\rho$ lead to inaccurate values of longitudinal acoustic waves speed, spanning from 5500 to 11000 m/s. A precise determination of these parameters is out of the scope of this work, however a quantitative analysis is required for the numerical simulations and further discussion.

\begin{figure}
	\par
	\begin{center}
		\includegraphics[scale=0.38]{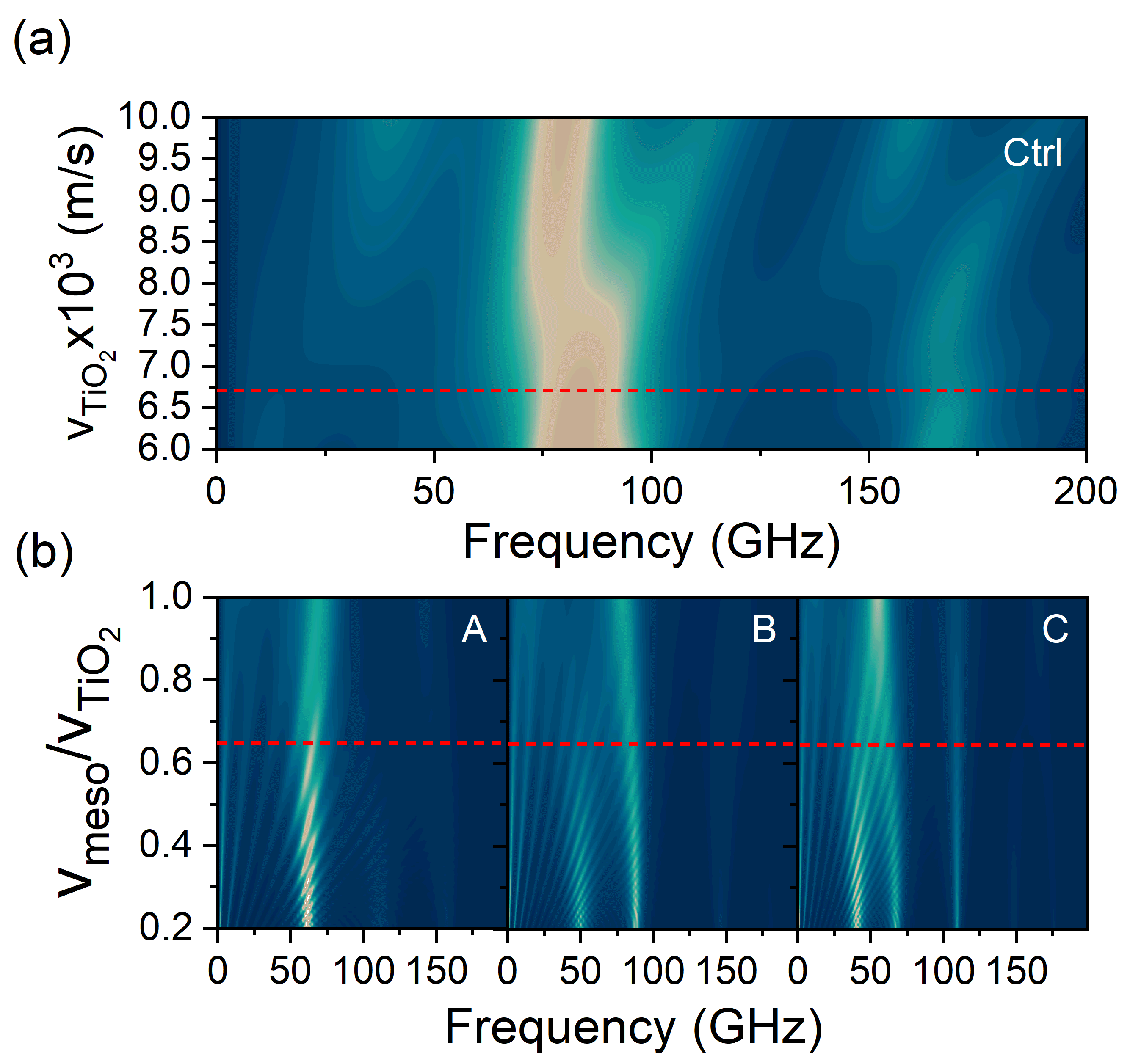} 
	\end{center}
	\par
	\vspace{-0.5cm} \caption{Colorplot of surface displacement amplitude vs frequency and (a)  dense TiO$_{2}$ speed of sound for control sample, and (b) speed of sound ratio between mesoporous and dense materials for mesoporous TiO$_{2}$-based samples. The dashed lines at V$_{TiO2} = 6700$ m/s and V$_{meso}/$V$_{TiO2} = 0.65$ correspond to the values that best match with experimental results.}
	\label{sim}
\end{figure}

We simulate the surface displacement at different TiO$_{2}$ sound speeds for the control sample, within the range of 6000 to 10000 m/s, and present the results on Fig.~\ref{sim}(a), in a colorplot. The spectrum that best matches with the experiment is simulated with v$_{TiO_{2}}=$ 6700 m/s, indicated by a dashed red line. The density used in the simulation is set constant at 2.9 g/cm$^{3}$. The TiO$_{2}$ dense layer parameters are then used as input for the calculation of the mesoporous samples.

The mesoporous TiO$_{2}$ layer parameters are extracted according to the layer porosity and the dense TiO$_{2}$ constants. The refractive index is calculated using the Bruggemann approximation of a complex composite by an effective homogeneous medium for a mixture of two dielectric media states.~\cite{fungApplicationValidityEffective2019} The mass density is considered to be the same as the dense layer,~\cite{Abdala2020} and the speed of sound in the mesoporous material is varied to match with the experimental results. The evolution of the surface displacement spectrum as a function of the ratio between the sound velocities of the mesoporous and dense TiO$_{2}$ layers for the three samples is shown on the three panels of Fig.~\ref{sim}(b). The dashed red lines indicate the ratio of the sound velocities for which the simulations best match with the experimental results ($v_{meso} = 4355$ m/s), and the respective spectra are displayed in Fig.~\ref{fft}(b). They exhibit a good agreement with the experiment after adjustments on the layers thicknesses according to the values from Table~\ref{table_1}, with a maximum deviation of 11\% from the nominal values for the MTFs. We also consider acoustic losses with an effective phonon decay length of $\sim 75$ nm.~\cite{Abdala2020,ortiz_phonon_2019}. Considering the values $v_{meso} = 4355$ m/s, $\rho=2.9$ g/cm$^{3}$ and the Poisson ratio for mesoporous materials $\nu=0.2$, we deduce that the Young's modulus of the thin films is $\sim 49.5$ GPa. Although this method of deducing Young's modulus is not standard, the estimated value is in close agreement with the ones reported in the literature, between 37 and 50 GPa.~\cite{lionello_mechanical_2022-1,lionello_structural_2017-1}

In Fig.~\ref{sim}(b) we observe a clear increase in mode frequency for the low-frequency modes upon rising the speed of sound in the mesoporous layer. In contrast, the high-frequency modes are hardly affected. This implies that the low-frequency resonances are indeed mainly confined within the MTF. It is worth mentioning that the studied variation of the sound velocity in the mesoporous layer can be experimentally achieved by liquid infiltration into the nanopores, which modifies the elastic properties of the effective medium composed of the TiO$_{2}$ dense matrix and air in the pores. As chemical adsorption and capillary condensation are reversible processes, MTFs can be employed as active elements in optoacoustic sensing devices.

The results exhibited in this section are compatible with what has been reported on SiO$_{2}$ mesoporous systems,~\cite{Abdala2020} and reinforce the feasibility of coherent acoustic phonon generation and detection in different mesoporous materials. Furthermore, this work reinforces the concept of mesoporous thin films as potential environment-responsive platforms able to transduce a physicochemical process (adsorption, capillary condensation) into optical~\cite{gazoni_designed_2017,auguie_tamm_2014,fuertes_photonic_2007} or nanoacoustic signals~\cite{Abdala2020}. This building block represents an important step towards the engineering of more complex structures based on such soft materials for practical applications, e.g., sensing.

\section{Conclusion}

In this work we have demonstrated the confinement of gigahertz coherent acoustic phonons in mesoporous titanium dioxide structures, extending the concept of mesoporous acoustic resonators to different material systems. Simulations on photoelastic interaction and surface displacement convey appreciable agreement with the experimental results. Furthermore, we discussed the effects of changes in the elastic parameters on the mesoporous resonator performance. These changes can be induced by chemical compound infiltration. This concept can be extended and applied to novel sensing devices based on ultrahigh-frequency resonators, with the mesoporous layer as the active optoacoustic element. Our findings unlock the way to a promising platform for nanoacoustic sensing and reconfigurable optoacoustic nanodevices based on soft, inexpensive fabrication methods. 

\section{Acknowledgment}
The authors acknowledge funding by the European Research Council Starting Grant No. 715939, Nanophennec, the ECOS-Sud Program through the project TUNA-Phon (PA19N03), and through a public grant overseen by the ANR as part of the “Investissements d’Avenir” program (Labex NanoSaclay Grant No. ANR-10-LABX-0035). GJAASI acknowledges ANPCyT for projects PICT 2017-4651, PICT-2018-04236 and PICT 2020-03130. MCF acknowledges ANPCyT for projects PICT 2015-0351 and 2017-1133.ME acknowledges funding by the University of Oldenburg through a Carl von Ossietzky Young Researchers' Fellowship. NDLK and GJAASI acknowledge funding from CNRS through the project IEA RANas. 

\section{Authors contributions}
GJAASI and NDLK proposed the concepts and directed the research. NLA, MCF, AB, HP, and GJAASI fabricated and characterized the samples.  BP and NDLK performed the picosecond ultrasonics experiments. ERCO, CX, ME, NLA, and NDLK performed the simulations and analyzed the data.

\bibliography{TiO2_mesoporous.bbl}

\begin{thebibliography}{10}

\bibitem{Trigo2002}
M.~Trigo, A.~Bruchhausen, A.~Fainstein, B.~Jusserand, and V.~Thierry-Mieg.
\newblock Confinement of acoustical vibrations in a semiconductor planar phonon
  cavity.
\newblock {\em Phys. Rev. Lett.}, 89:227402, Nov 2002.

\bibitem{balandinNanophononicsPhononEngineering2005}
Alexander~A. Balandin.
\newblock Nanophononics: {{Phonon Engineering}} in {{Nanostructures}} and
  {{Nanodevices}}.
\newblock {\em Journal of Nanoscience and Nanotechnology}, 5(7):1015--1022,
  July 2005.

\bibitem{rozas_lifetime_2009}
G.~Rozas, M.~F.~Pascual Winter, B.~Jusserand, A.~Fainstein, B.~Perrin,
  E.~Semenova, and A.~Lema{\^i}tre.
\newblock Lifetime of {{THz Acoustic Nanocavity Modes}}.
\newblock {\em Physical Review Letters}, 102(1):015502, January 2009.

\bibitem{beardsley_coherent_2010}
R.~P. Beardsley, A.~V. Akimov, M.~Henini, and A.~J. Kent.
\newblock Coherent {{Terahertz Sound Amplification}} and {{Spectral Line
  Narrowing}} in a {{Stark Ladder Superlattice}}.
\newblock {\em Physical Review Letters}, 104(8):085501, February 2010.

\bibitem{Volz2016}
Sebastian Volz, Jose Ordonez-Miranda, Mika~Prunnila Andrey~Shchepetov, Jouni
  Ahopelto, Thomas Pezeril, Gwenaelle Vaudel, Vitaly Gusev, Pascal Ruello, Eva
  M.Weig, Martin Schubert, Mike Hettich, Martin Grossman, Thomas Dekorsy,
  Francesc Alzina, Bartlomiej Graczykowski, Emigdio Chavez-Angel, J.~Sebastian
  Reparaz, Markus~R. Wagner, Clivia~M. Sotomayor-Torres, Shiyun Xiong,
  Sanghamitra Neogi, and Davide Donadio.
\newblock {Nanophononics: State of the art and perspectives}.
\newblock {\em European Physical Journal B}, 89(1):15, 2016.

\bibitem{della_picca_tailored_2016}
Fabricio Della~Picca, Rodrigo Berte, Mohsen Rahmani, Pablo Albella, Juan~M.
  Bujjamer, Mart{\'i}n Poblet, Emiliano Cort{\'e}s, Stefan~A. Maier, and
  Andrea~V. Bragas.
\newblock Tailored {{Hypersound Generation}} in {{Single Plasmonic
  Nanoantennas}}.
\newblock {\em Nano Letters}, 16(2):1428--1434, February 2016.

\bibitem{lamberti_optomechanical_2017}
F.~R. Lamberti, Q.~Yao, L.~Lanco, D.~T. Nguyen, M.~Esmann, A.~Fainstein,
  P.~Sesin, S.~Anguiano, V.~Villafa{\~n}e, A.~Bruchhausen, P.~Senellart,
  I.~Favero, and N.~D. {Lanzillotti-Kimura}.
\newblock Optomechanical properties of {{GaAs}}/{{AlAs}} micropillar resonators
  operating in the 18 {{GHz}} range.
\newblock {\em Optics Express}, 25(20):24437, October 2017.

\bibitem{de_luca_phonon_2019}
Marta De~Luca, Claudia Fasolato, Marcel~A. Verheijen, Yizhen Ren, Milo~Y.
  Swinkels, Sebastian K{\"o}lling, Erik P. A.~M. Bakkers, Riccardo Rurali,
  Xavier Cartoix{\`a}, and Ilaria Zardo.
\newblock Phonon {{Engineering}} in {{Twinning Superlattice Nanowires}}.
\newblock {\em Nano Letters}, 19(7):4702--4711, July 2019.

\bibitem{esmann_brillouin_2019}
M.~Esmann, F.~R. Lamberti, A.~Harouri, L.~Lanco, I.~Sagnes, I.~Favero,
  G.~Aubin, C.~{Gomez-Carbonell}, A.~Lema{\^i}tre, O.~Krebs, P.~Senellart, and
  N.~D. {Lanzillotti-Kimura}.
\newblock Brillouin scattering in hybrid optophononic {{Bragg}} micropillar
  resonators at 300 {{GHz}}.
\newblock {\em Optica}, 6(7):854, July 2019.

\bibitem{arbouet_optical_2006}
A.~Arbouet, N.~Del~Fatti, and F.~Vallee.
\newblock Optical control of the coherent acoustic vibration of metal
  nanoparticles.
\newblock {\em The Journal of Chemical Physics}, 124(14):144701, April 2006.

\bibitem{obrien_ultrafast_2014}
Kevin O'Brien, N.~D. {Lanzillotti-Kimura}, Junsuk Rho, Haim Suchowski, Xiaobo
  Yin, and Xiang Zhang.
\newblock Ultrafast acousto-plasmonic control and sensing in complex
  nanostructures.
\newblock {\em Nature Communications}, 5(1):4042, September 2014.

\bibitem{guillet_ultrafast_2019}
Yannick Guillet, Allaoua Abbas, Serge Ravaine, and Bertrand Audoin.
\newblock Ultrafast microscopy of the vibrational landscape of a single
  nanoparticle.
\newblock {\em Applied Physics Letters}, 114(9):091904, March 2019.

\bibitem{poblet_acoustic_2021}
Martin Poblet, Rodrigo Bert{\'e}, Hilario~D. Boggiano, Yi~Li, Emiliano
  Cort{\'e}s, Gustavo Grinblat, Stefan~A. Maier, and Andrea~V. Bragas.
\newblock Acoustic {{Coupling}} between {{Plasmonic Nanoantennas}}:
  {{Detection}} and {{Directionality}} of {{Surface Acoustic Waves}}.
\newblock {\em ACS Photonics}, 8(10):2846--2852, October 2021.

\bibitem{lanzillotti-kimura_polarization-controlled_2018}
Norberto~D. {Lanzillotti-Kimura}, Kevin~P. O'Brien, Junsuk Rho, Haim Suchowski,
  Xiaobo Yin, and Xiang Zhang.
\newblock Polarization-controlled coherent phonon generation in
  acoustoplasmonic metasurfaces.
\newblock {\em Physical Review B}, 97(23):235403, June 2018.

\bibitem{soukiassian_acoustic_2007}
A.~Soukiassian, W.~Tian, D.~A. Tenne, X.~X. Xi, D.~G. Schlom, N.~D.
  {Lanzillotti-Kimura}, A.~Bruchhausen, A.~Fainstein, H.~P. Sun, X.~Q. Pan,
  A.~Cros, and A.~Cantarero.
\newblock Acoustic {{Bragg}} mirrors and cavities made using piezoelectric
  oxides.
\newblock {\em Applied Physics Letters}, 90(4):042909, January 2007.

\bibitem{lanzillotti-kimura_enhancement_2010}
N.~D. {Lanzillotti-Kimura}, A.~Fainstein, B.~Perrin, B.~Jusserand,
  A.~Soukiassian, X.~X. Xi, and D.~G. Schlom.
\newblock Enhancement and {{Inhibition}} of {{Coherent Phonon Emission}} of a
  {{Ni Film}} in a {{BaTiO}} 3 / {{SrTiO}} 3 {{Cavity}}.
\newblock {\em Physical Review Letters}, 104(18):187402, May 2010.

\bibitem{vasileiadis_progress_2021}
Thomas Vasileiadis, Jeena Varghese, Visnja Babacic, Jordi {Gomis-Bresco},
  Daniel Navarro~Urrios, and Bartlomiej Graczykowski.
\newblock Progress and perspectives on phononic crystals.
\newblock {\em Journal of Applied Physics}, 129(16):160901, April 2021.

\bibitem{cang_fundamentals_2022}
Yu~Cang, Yabin Jin, Bahram {Djafari-Rouhani}, and George Fytas.
\newblock Fundamentals, progress and perspectives on high-frequency phononic
  crystals.
\newblock {\em Journal of Physics D: Applied Physics}, 55(19):193002, May 2022.

\bibitem{lanzillotti-kimura_resonant_2009}
N.~D. {Lanzillotti-Kimura}, A.~Fainstein, B.~Jusserand, and A.~Lema{\^i}tre.
\newblock Resonant {{Raman}} scattering of nanocavity-confined acoustic
  phonons.
\newblock {\em Physical Review B}, 79(3):035404, January 2009.

\bibitem{anguianoMicropillarResonatorsOptomechanics2017}
S.~Anguiano, A.~E. Bruchhausen, B.~Jusserand, I.~Favero, F.~R. Lamberti,
  L.~Lanco, I.~Sagnes, A.~Lema{\^i}tre, N.~D. {Lanzillotti-Kimura},
  P.~Senellart, and A.~Fainstein.
\newblock Micropillar {{Resonators}} for {{Optomechanics}} in the {{Extremely
  High}} 19\textendash 95-{{GHz Frequency Range}}.
\newblock {\em Physical Review Letters}, 118(26):263901, June 2017.

\bibitem{chafatinos_polariton-driven_2020}
D.~L. Chafatinos, A.~S. Kuznetsov, S.~Anguiano, A.~E. Bruchhausen, A.~A.
  Reynoso, K.~Biermann, P.~V. Santos, and A.~Fainstein.
\newblock Polariton-driven phonon laser.
\newblock {\em Nature Communications}, 11(1):4552, December 2020.

\bibitem{ortiz_topological_2021}
O.~Ortiz, P.~Priya, A.~Rodriguez, A.~Lemaitre, M.~Esmann, and N.~D.
  {Lanzillotti-Kimura}.
\newblock Topological optical and phononic interface mode by simultaneous band
  inversion.
\newblock {\em Optica}, 8(5):598, May 2021.

\bibitem{arregui_coherent_2019}
G.~Arregui, O.~Ort{\'i}z, M.~Esmann, C.~M. {Sotomayor-Torres},
  C.~{Gomez-Carbonell}, O.~Mauguin, B.~Perrin, A.~Lema{\^i}tre, P.~D.
  Garc{\'i}a, and N.~D. {Lanzillotti-Kimura}.
\newblock Coherent generation and detection of acoustic phonons in topological
  nanocavities.
\newblock {\em APL Photonics}, 4(3):030805, March 2019.

\bibitem{lanzillotti-kimura_enhanced_2011}
N.~D. {Lanzillotti-Kimura}, A.~Fainstein, B.~Perrin, B.~Jusserand, L.~Largeau,
  O.~Mauguin, and A.~Lemaitre.
\newblock Enhanced optical generation and detection of acoustic nanowaves in
  microcavities.
\newblock {\em Physical Review B}, 83(20):201103, May 2011.

\bibitem{soler-illia_mesoporous_2006}
G.~J. A.~A. {Soler-Illia} and P.~Innocenzi.
\newblock Mesoporous {{Hybrid Thin Films}}: {{The Physics}} and {{Chemistry
  Beneath}}.
\newblock {\em Chemistry - A European Journal}, 12(17):4478--4494, June 2006.

\bibitem{gazoni_designed_2017}
Rodrigo~Mart{\'i}nez Gazoni, Mart{\'i}n~G. Bellino, M.~Cecilia~Fuertes, Gustavo
  Gim{\'e}nez, Galo J. A.~A. {Soler-Illia}, and Mar{\'i}a~Luz {Mart{\'i}nez
  Ricci}.
\newblock Designed nanoparticle\textendash mesoporous multilayer nanocomposites
  as tunable plasmonic\textendash photonic architectures for electromagnetic
  field enhancement.
\newblock {\em Journal of Materials Chemistry C}, 5(14):3445--3455, 2017.

\bibitem{gomopoulos_one-dimensional_2010}
N.~Gomopoulos, D.~Maschke, C.~Y. Koh, E.~L. Thomas, W.~Tremel, H.-J. Butt, and
  G.~Fytas.
\newblock One-{{Dimensional Hypersonic Phononic Crystals}}.
\newblock {\em Nano Letters}, 10(3):980--984, March 2010.

\bibitem{Benetti2018}
Giulio Benetti, Marco Gandolfi, Margriet~J. Van~Bael, Luca Gavioli, Claudio
  Giannetti, Claudia Caddeo, and Francesco Banfi.
\newblock Photoacoustic sensing of trapped fluids in nanoporous thin films:
  Device engineering and sensing scheme.
\newblock {\em ACS Applied Materials \& Interfaces}, 10(33):27947--27954, 2018.
\newblock PMID: 30039696.

\bibitem{Abdala2020}
Nicolas~Lopez Abdala, Martin Esmann, Maria~C. Fuertes, Paula~C. Angelomé, Omar
  Ortiz, Axel Bruchhausen, Hernan Pastoriza, Bernard Perrin, Galo J. A.~A.
  Soler-Illia, and Norberto~D. Lanzillotti-Kimura.
\newblock Mesoporous thin films for acoustic devices in the gigahertz range.
\newblock {\em The Journal of Physical Chemistry C}, 124(31):17165--17171,
  2020.

\bibitem{schneider_engineering_2012}
Dirk Schneider, Faroha Liaqat, El~Houssaine El~Boudouti, Youssef El~Hassouani,
  Bahram {Djafari-Rouhani}, Wolfgang Tremel, Hans-J{\"u}rgen Butt, and George
  Fytas.
\newblock Engineering the {{Hypersonic Phononic Band Gap}} of {{Hybrid Bragg
  Stacks}}.
\newblock {\em Nano Letters}, 12(6):3101--3108, June 2012.

\bibitem{schneider_defect-controlled_2013}
Dirk Schneider, Faroha Liaqat, El~Houssaine El~Boudouti, Ossama El~Abouti,
  Wolfgang Tremel, Hans-J{\"u}rgen Butt, Bahram {Djafari-Rouhani}, and George
  Fytas.
\newblock Defect-{{Controlled Hypersound Propagation}} in {{Hybrid
  Superlattices}}.
\newblock {\em Physical Review Letters}, 111(16):164301, October 2013.

\bibitem{auguie_tamm_2014}
Baptiste Augui{\'e}, Mar{\'i}a~Cecilia Fuertes, Paula~C. Angelom{\'e},
  Nicol{\'a}s~L{\'o}pez Abdala, Galo J. A.~A. Soler~Illia, and Alejandro
  Fainstein.
\newblock Tamm {{Plasmon Resonance}} in {{Mesoporous Multilayers}}: {{Toward}}
  a {{Sensing Application}}.
\newblock {\em ACS Photonics}, 1(9):775--780, September 2014.

\bibitem{fuertes_photonic_2007}
M.~C. Fuertes, F.~J. {L{\'o}pez-Alcaraz}, M.~C. Marchi, H.~E. Troiani, V.~Luca,
  H.~M{\'i}guez, and G.~J. A.~A. {Soler-Illia}.
\newblock Photonic {{Crystals}} from {{Ordered Mesoporous Thin-Film Functional
  Building Blocks}}.
\newblock {\em Advanced Functional Materials}, 17(8):1247--1254, May 2007.

\bibitem{thelen_laser-excited_2021}
Marc Thelen, Nicolas Bochud, Manuel Brinker, Claire Prada, and Patrick Huber.
\newblock Laser-excited elastic guided waves reveal the complex mechanics of
  nanoporous silicon.
\newblock {\em Nature Communications}, 12(1):3597, June 2021.

\bibitem{suarez_photocatalytic_2011}
S.~Su{\'a}rez, N.~Arconada, Y.~Castro, J.M. Coronado, R.~Portela, A.~Dur{\'a}n,
  and B.~S{\'a}nchez.
\newblock Photocatalytic degradation of {{TCE}} in dry and wet air conditions
  with {{TiO2}} porous thin films.
\newblock {\em Applied Catalysis B: Environmental}, 108--109:14--21, October
  2011.

\bibitem{hussein_mesoporous_2013}
Abdulmenan~M. Hussein, Luther Mahoney, Rui Peng, Harrison Kibombo, Chia-Ming
  Wu, Ranjit~T. Koodali, and Rajesh Shende.
\newblock Mesoporous coupled {{ZnO}}/{{TiO}} {\textsubscript{2}} photocatalyst
  nanocomposites for hydrogen generation.
\newblock {\em Journal of Renewable and Sustainable Energy}, 5(3):033118, May
  2013.

\bibitem{harmankaya_raloxifene_2013}
N.~Harmankaya, J.~Karlsson, A.~Palmquist, M.~Halvarsson, K.~Igawa,
  M.~Andersson, and P.~Tengvall.
\newblock Raloxifene and alendronate containing thin mesoporous titanium oxide
  films improve implant fixation to bone.
\newblock {\em Acta Biomaterialia}, 9(6):7064--7073, June 2013.

\bibitem{violi_highly_2012}
Ianina~L. Violi, M.~Dolores Perez, M.~Cecilia Fuertes, and Galo J. A.~A.
  {Soler-Illia}.
\newblock Highly {{Ordered}}, {{Accessible}} and {{Nanocrystalline Mesoporous
  TiO}} {\textsubscript{2}} {{Thin Films}} on {{Transparent Conductive
  Substrates}}.
\newblock {\em ACS Applied Materials \& Interfaces}, 4(8):4320--4330, August
  2012.

\bibitem{pan_block_2011}
Jia~Hong Pan, X.S. Zhao, and Wan~In Lee.
\newblock Block copolymer-templated synthesis of highly organized mesoporous
  {{TiO2-based}} films and their photoelectrochemical applications.
\newblock {\em Chemical Engineering Journal}, 170(2-3):363--380, June 2011.

\bibitem{baldini_phonon-driven_2018}
Edoardo Baldini, Tania Palmieri, Adriel Dominguez, Pascal Ruello, Angel Rubio,
  and Majed Chergui.
\newblock Phonon-{{Driven Selective Modulation}} of {{Exciton Oscillator
  Strengths}} in {{Anatase TiO}} {\textsubscript{2}} {{Nanoparticles}}.
\newblock {\em Nano Letters}, 18(8):5007--5014, August 2018.

\bibitem{baldini_exciton_2019}
Edoardo Baldini, Adriel Dominguez, Tania Palmieri, Oliviero Cannelli, Angel
  Rubio, Pascal Ruello, and Majed Chergui.
\newblock Exciton control in a room temperature bulk semiconductor with
  coherent strain pulses.
\newblock {\em Science Advances}, 5(11):eaax2937, November 2019.

\bibitem{perrin_interferometric_1999}
B~Perrin, C~Rossignol, B~Bonello, and J.-C Jeannet.
\newblock Interferometric detection in picosecond ultrasonics.
\newblock {\em Physica B: Condensed Matter}, 263--264:571--573, March 1999.

\bibitem{matsuda_reflection_2002-1}
O.~Matsuda and O.~B. Wright.
\newblock Reflection and transmission of light in multilayers perturbed by
  picosecond strain pulse propagation.
\newblock {\em Journal of the Optical Society of America B}, 19(12):3028,
  December 2002.

\bibitem{angelome_multifunctional_2006}
P.~C. Angelom{\'e}, M.~C. Fuertes, and G.~J. A.~A. {Soler-Illia}.
\newblock Multifunctional, {{Multilayer}}, {{Multiscale}}: {{Integrative
  Synthesis}} of {{Complex Macroporous}} and {{Mesoporous Thin Films}} with
  {{Spatial Separation}} of {{Porosity}} and {{Function}}.
\newblock {\em Advanced Materials}, 18(18):2397--2402, September 2006.

\bibitem{fuertes_sorption_2008}
Mar{\'i}a~Cecilia Fuertes, Silvia Colodrero, Gabriel Lozano, Agust{\'i}n~R.
  {Gonz{\'a}lez-Elipe}, David Grosso, C{\'e}dric Boissi{\`e}re, Cl{\'e}ment
  S{\'a}nchez, Galo J. de A.~A. {Soler-Illia}, and Hern{\'a}n M{\'i}guez.
\newblock Sorption {{Properties}} of {{Mesoporous Multilayer Thin Films}}.
\newblock {\em The Journal of Physical Chemistry C}, 112(9):3157--3163, March
  2008.

\bibitem{soler-illia_critical_2012}
Galo J. A.~A. {Soler-Illia}, Paula~C. Angelom{\'e}, M.~Cecilia Fuertes, David
  Grosso, and Cedric Boissiere.
\newblock Critical aspects in the production of periodically ordered mesoporous
  titania thin films.
\newblock {\em Nanoscale}, 4(8):2549, 2012.

\bibitem{huynh_subterahertz_2006}
A.~Huynh, N.~D. {Lanzillotti-Kimura}, B.~Jusserand, B.~Perrin, A.~Fainstein,
  M.~F. {Pascual-Winter}, E.~Peronne, and A.~Lema{\^i}tre.
\newblock Subterahertz {{Phonon Dynamics}} in {{Acoustic Nanocavities}}.
\newblock {\em Physical Review Letters}, 97(11):115502, September 2006.

\bibitem{boissiere_porosity_2005}
C{\'e}dric Boissiere, David Grosso, Sophie Lepoutre, Lionel Nicole,
  Aline~Brunet Bruneau, and Cl{\'e}ment Sanchez.
\newblock Porosity and {{Mechanical Properties}} of {{Mesoporous Thin Films
  Assessed}} by {{Environmental Ellipsometric Porosimetry}}.
\newblock {\em Langmuir}, 21(26):12362--12371, December 2005.

\bibitem{ruello_physical_2015}
Pascal Ruello and Vitalyi~E. Gusev.
\newblock Physical mechanisms of coherent acoustic phonons generation by
  ultrafast laser action.
\newblock {\em Ultrasonics}, 56:21--35, February 2015.

\bibitem{thomsen_surface_1986}
C.~Thomsen, H.~T. Grahn, H.~J. Maris, and J.~Tauc.
\newblock Surface generation and detection of phonons by picosecond light
  pulses.
\newblock {\em Physical Review B}, 34(6):4129--4138, September 1986.

\bibitem{Fainstein2007}
Alejandro Fainstein and Bernard Jusserand.
\newblock Raman scattering in resonant cavities.
\newblock In Manuel Cardona and Roberto Merlin, editors, {\em Light Scattering
  in Solid {{IX}}}, pages 17--110. {Springer Berlin Heidelberg}, {Berlin,
  Heidelberg}, 2007.

\bibitem{lanzillotti-kimura_theory_2011-1}
N.~D. {Lanzillotti-Kimura}, A.~Fainstein, B.~Perrin, and B.~Jusserand.
\newblock Theory of coherent generation and detection of {{THz}} acoustic
  phonons using optical microcavities.
\newblock {\em Physical Review B}, 84(6):064307, August 2011.

\bibitem{lanzillotti-kimura_coherent_2011-1}
N.~D. {Lanzillotti-Kimura}, A.~Fainstein, A.~Lemaitre, B.~Jusserand, and
  B.~Perrin.
\newblock Coherent control of sub-terahertz confined acoustic nanowaves:
  {{Theory}} and experiments.
\newblock {\em Physical Review B}, 84(11):115453, September 2011.

\bibitem{fungApplicationValidityEffective2019}
Tsun~Hang Fung, Tom Veeken, David Payne, Binesh Veettil, Albert Polman, and
  Malcolm Abbott.
\newblock Application and validity of the effective medium approximation to the
  optical properties of nano-textured silicon coated with a dielectric layer.
\newblock {\em Optics Express}, 27(26):38645, December 2019.

\bibitem{pillai_sol-gel_2017}
Sanjay~Gopal Ullattil and Pradeepan Periyat.
\newblock Sol-{{Gel Synthesis}} of {{Titanium Dioxide}}.
\newblock In Suresh~C. Pillai and Sarah Hehir, editors, {\em Sol-{{Gel
  Materials}} for {{Energy}}, {{Environment}} and {{Electronic Applications}}},
  pages 271--283. {Springer International Publishing}, {Cham}, 2017.

\bibitem{yang_compositional_2005}
Li-Lan Yang, Yi-Sheng Lai, J.S. Chen, P.H. Tsai, C.L. Chen, and C.~Jason Chang.
\newblock Compositional {{Tailored Sol-Gel SiO}} {\textsubscript{2}}
  \textendash{{TiO}} {\textsubscript{2}} {{Thin Films}}: {{Crystallization}},
  {{Chemical Bonding Configuration}}, and {{Optical Properties}}.
\newblock {\em Journal of Materials Research}, 20(11):3141--3149, November
  2005.

\bibitem{ottermann_youngs_1996}
C.~R. Ottermann, R.~Kuschnereit, O.~Anderson, P.~Hess, and K.~Bange.
\newblock Young's {{Modulus}} and {{Density}} of thin {{TiO}}
  {\textsubscript{2}} {{Films Produced}} by {{Different Methods}}.
\newblock {\em MRS Proceedings}, 436:251, 1996.

\bibitem{ferrara_hydrophilic_2010}
M~C Ferrara, L~Pilloni, S~Mazzarelli, and L~Tapfer.
\newblock Hydrophilic and optical properties of nanostructured titania prepared
  by sol\textendash gel dip coating.
\newblock {\em Journal of Physics D: Applied Physics}, 43(9):095301, March
  2010.

\bibitem{olofinjana_evaluation_2000}
A.O Olofinjana, J.M Bell, and A.K J{\"a}mting.
\newblock Evaluation of the mechanical properties of sol\textendash
  gel-deposited titania films using ultra-micro-indentation method.
\newblock {\em Wear}, 241(2):174--179, July 2000.

\bibitem{bendavid_deposition_2000}
A~Bendavid, P~J Martin, and H~Takikawa.
\newblock Deposition and modification of titanium dioxide thin
  films by filtered arc deposition.
\newblock {\em Thin Solid Films}, page~9, 2000.

\bibitem{soares_hardness_2008}
Paulo Soares, Alexandre Mikowski, Carlos~M. Lepienski, Emanuel Santos,
  Gl{\'o}ria~A. Soares, Vitoldo~Swinka Filho, and Neide~K. Kuromoto.
\newblock Hardness and elastic modulus of {{TiO2}} anodic films measured by
  instrumented indentation.
\newblock {\em Journal of Biomedical Materials Research Part B: Applied
  Biomaterials}, 84B(2):524--530, February 2008.

\bibitem{borgese_young_2012}
L.~Borgese, M.~Gelfi, E.~Bontempi, P.~Goudeau, G.~Geandier, D.~Thiaudi{\`e}re,
  and L.E. Depero.
\newblock Young modulus and {{Poisson}} ratio measurements of {{TiO2}} thin
  films deposited with {{Atomic Layer Deposition}}.
\newblock {\em Surface and Coatings Technology}, 206(8-9):2459--2463, January
  2012.

\bibitem{ortiz_phonon_2019}
O.~Ort{\'i}z, M.~Esmann, and N.~D. {Lanzillotti-Kimura}.
\newblock Phonon engineering with superlattices: {{Generalized}} nanomechanical
  potentials.
\newblock {\em Physical Review B}, 100(8):085430, August 2019.

\bibitem{lionello_mechanical_2022-1}
Diego~F. Lionello, Juan~Ignacio Ramallo, Galo J. A.~A. {Soler-Illia}, and
  Mar{\'i}a~Cecilia Fuertes.
\newblock Mechanical properties of ordered mesoporous oxides thin films.
\newblock {\em Journal of Sol-Gel Science and Technology}, 101(1):114--139,
  January 2022.

\bibitem{lionello_structural_2017-1}
Diego~F. Lionello, Paula~Y. Steinberg, M.~Mercedes Zalduendo, Galo J. A.~A.
  {Soler-Illia}, Paula~C. Angelom{\'e}, and M.~Cecilia Fuertes.
\newblock Structural and {{Mechanical Evolution}} of {{Mesoporous Films}} with
  {{Thermal Treatment}}: {{The Case}} of {{Brij}} 58 {{Templated Titania}}.
\newblock {\em The Journal of Physical Chemistry C}, 121(40):22576--22586,
  October 2017.

\end{thebibliography}

\end{document}